# Two Electrons in a Quantum Dot: A Unified Approach

Bülent Gönül, Ebru Bakır and Koray Köksal

Department of Engineering Physics, University of Gaziantep, 27310, Gaziantep-Turkey

**Abstract**

Low-lying energy levels of two interacting electrons confined in a two-dimensional parabolic quantum dot in the presence of an external magnetic field have been revised within the frame of a novel model. The present formalism, which gives closed algebraic solutions for the specific values of magnetic field and spatial confinement length, enables us to see explicitly individual effects of the electron correlation.

## 1. INTRODUCTION

Two-dimensional hydrogen atom in a magnetic field has been subject of active research during the last years [1]. This problem is of practical interest because of the technological advances in nanofabrication technology that have made it possible to create low-dimensional structures like quantum wells, quantum wires and quantum dots. A large body of articles has been published on this problem in the framework of non-relativistic quantum mechanics as relativistic effects are not considerably significant in semiconductor devices.

In particular, quantum dots in which only a few electrons are bound at semiconductor interfaces have been the subject of intense research studies over the last few years. The electron motion in quantum dots are confined to a region with dimensions comparable to the de Broglie wavelength of the particle. The result is the quantization of energy. However, since the quantization in the vertical direction is much stronger than in the planar directions, a quantum dot can well be treated as a two dimensional disc of finite radius. The intensive investigations have revealed that optoelectronic properties of such systems are quite sensitive to the reduction of their dimensionality and to the strength of applied external magnetic field, and depend strongly on the electron-electron interaction. Different methods have been used in the related literature to search the energy spectrum and the correlation effects of the interacting electrons in such systems. For most recent reviews, see Ref. [2, 3].

However, an exact solution of the Schrödinger equation for a many-electron system is not possible in general. Thus, very little is known about the nature of the electron correlation even in simple systems. Nevertheless, insight into the correlation problem can be obtained through the study of exactly solvable model systems in some specific cases. For this purpose it is simplest to consider a system of two particles bound by a suitable central potential for which an exact solution is possible. This simplification, at least in part, provides scope for much further studies.

Therefore, in the recent literature regarding quantum dots attention mainly has been focused on understanding the quantum mechanical behaviour of two interacting electrons confined in various two dimensional dot geometries under the influence of an external magnetic field, due to the fact that electron-electron interactions which are known to be quite important in such quasi-zero dimensional structures are enhanced by the presence an additional confinement arising from the magnetic field. However, the complicated nature of the recursion relations appeared in solving the associated radial part of the relevant Schrödinger equation of even this simplified interacting two-electron case does not allow in general an exact solution, except for the case that some certain relations between the Coulomb repulsion strength and the strength of the magnetic field and/or spatial confinement exist. As a result, the studies on exact treatments so far have been content with just obtaining a few eigenvalues and their related eigenstates.

Within this context, using the spirit of the novel approximation proposed more recently [4], we suggest here an alternative scheme for the treatment of the problem interested, which capables of determining the general closed form solution for such states, in terms of associated Laguerre polynomials, together with corresponding eigenvalues. The proposed algebraic structure allows one to see clearly the physics behind individual contributions in constructing analytical solutions. A full description of the quasi-exact solutions to the problem, together with the comparative analysis of the analytical results obtained within the frame of the present formalism with those previously found in the literature, are given. Although a similar idea to provide closed-form expressions for the solutions of the same problem has recently appeared [2], the prescription suggested in this article decomposes elegantly the related Schrödinger equation involving a quasi-exactly solvable potential, which is not feasible in [2], in order to comprehend how two-body correlation affects the exactly solvable piece of the entire problem. To our knowledge, the literature does not cover such an investigation.

The paper is organized as follows. In Section 2 we first give a theoretical background on the problem considered. The main idea of our approach is then summarized in the same section. The application of the present model to the problem leading to simple relations for the calculations at each successive orders of the modification function and the results obtained are shown in Section 3. The paper ends with a brief summary and concluding remarks.

## 2. THEORETICAL CONSIDERATIONS

### 2.1. Background of the problem

The Hamiltonian for a system of two interacting electrons in the presence of both an external uniform magnetic field, applied along $z$ – axis, and a parabolic potential can be separated into the center-of-mass and relative motion parts as follows:

$$H_{cm} = \frac{1}{2M}\left(\bar{P} + \frac{Q}{c}\bar{A}(\bar{R})\right)^2 + \frac{1}{2}Mw_0^2\bar{R}^2 \tag{1}$$

$$H_{rm} = \frac{1}{2\mu}\left(\bar{p} + \frac{q}{c}\bar{A}(\bar{r})\right)^2 + \frac{1}{2}\mu w_0^2 r^2 + \frac{e^2}{\varepsilon}\frac{1}{|\bar{r}|} \tag{2}$$

where $\bar{R} = \frac{1}{2}(\bar{r}_1 + \bar{r}_2)$ and $\bar{P} = (\bar{p}_1 + \bar{p}_2)$ are the center-of-masss coordinates while $\bar{r} = \bar{r}_1 - \bar{r}_2$ and $\bar{p} = \frac{1}{2}(\bar{p}_1 - \bar{p}_2)$ are the relative coordinates, similarly $M = 2m^*$ and $Q = 2e$ are the total mass and charge, respectively, in the center-of-mass consideration while $\mu = m^*/2$ and $q = e/2$ are the reduced mass and charge in the relative motion system, and finally $w_0$ is the characteristic confinement frequency. Obviously, $m^*$ is the effective mass of each electron and $\varepsilon$ denotes the dielectric constant of the medium. Due to this separability, the wave function of the system considered reads simply as $\Psi(\bar{r},\bar{R}) = \Phi(\bar{r})\Theta(\bar{R})$, and the Schrödinger equation splits into two independent equations. Naturally, the total energy of the system in this case is $E = E_{cm} + E_{rm}$.

If the symmetric gauge $\bar{A}(\bar{R}) = \bar{B} \times \bar{R}/2$ is chosen for the vector potential of the magnetic field, Eq. (1) can then be written as a sum of two terms; the usual two dimensional isotropic harmonic oscillator Hamiltonian with frequency $w = \left[w_0^2 + \left(w_c^2/4\right)\right]^{1/2}$ plus a term proportional to $L_z$. Here, $w_c = QB/Mc$ is the cyclotron frequency and $L_z$ is the $z$-component of the angular momentum operator which commutes with the first part of the Hamiltonian.

As the solutions of Eq. (1) are well known in the literature, which can also be extracted easily from the present calculations, we mainly focus here on the relative motion Hamiltonian in Eq. (2) that is reduced to

$$H_{rm} = \frac{p^2}{2\mu} + \frac{1}{2}\mu w^2 r^2 + \frac{e^2}{\varepsilon}\frac{1}{|\vec{r}|} + \frac{w_c}{2}L_z \qquad (3)$$

Since we are dealing with a two-dimensional problem we choose to work in polar coordinates $(r,\varphi)$, consequently we can introduce the following ansatz for the eigenfunction

$$\Phi(\vec{r}) = r^{-1/2}\psi(r)e^{im\varphi}/\sqrt{2\pi} \qquad (4)$$

Substituting (4) into (3) one can readily obtain that the radial function $\psi(r)$ satisfies the second-order differential equations

$$\left\{-\frac{\hbar^2}{2\mu}\left[\frac{d^2}{dr^2} - \left(m^2 - \frac{1}{4}\right)\frac{1}{r^2}\right] + \frac{1}{2}\mu w^2 r^2 + \frac{e^2}{\varepsilon r} + \frac{w_c}{2}m\hbar - E_{rm}\right\}\psi(r) = 0 \qquad (5)$$

where $m = 0, \pm 1, \pm 2, \ldots$ is the azimuthal quantum number.

## 2.2. Formalism

So far many models have been introduced to construct analytical solutions of the above Hamiltonian. Bearing all these works in mind, we suggest here an alternative scheme for the treatment of such problems, leading to explicit understanding of the individual contributions coming from the distinct interaction terms appearing in (5), unlike the other models.

Let us first concentrate on the original form of the formalism [4], which has been developed in the light of a remarkable work [5] and employed successfully to discuss two significant problems in physics [6]. At this stage we note that this model presented below will eventually be improved in the next section for its proper applicability to the present problem. This would clarify the flexibility, consequently, the power of the formalism in treating exactly- and quasi-exactly solvable systems within the same frame.

Starting with the general form of the Schrödinger equation

$$\frac{\psi''(r)}{\psi(r)} = \frac{2\mu}{\hbar^2}[V(r) - E] \tag{6}$$

and remembering that its exact solutions usually take the form

$$\psi(r) = f(r)F[s(r)] \tag{7}$$

the substitution of (7) into (6) yields obviously the second-order differential equation

$$\left(\frac{f''}{f} + \frac{F''s'^2}{F} + \frac{s''F'}{F} + 2\frac{F's'f'}{Ff}\right) = \frac{2\mu}{\hbar^2}(V - E) \tag{8}$$

that is reduced to the form of well known hypergeometric (or confluent hypergeometric) equations

$$F'' + \left(\frac{s''}{s'^2} + 2\frac{f'}{s'f}\right)F' + \left[\frac{f''}{s'^2 f} + \frac{2\mu}{\hbar^2}\left(\frac{E - V}{s'^2}\right)\right]F = 0 \tag{9}$$

Reminding the general form of the differential equations, which reproduce closed analytical solutions through orthogonal polynomials,

$$F''(s) + \frac{\tau(s)}{\sigma(s)}F'(s) + \frac{\tilde{\sigma}(s)}{\sigma^2(s)}F(s) = 0 \tag{10}$$

where the forms of $\tau/\sigma$ and $\tilde{\sigma}/\sigma^2$ are well defined [7] for any special function $F(s)$, one obtains

$$\frac{s''}{s'^2} + 2\frac{f'}{s'f} = \frac{\tau}{\sigma} \quad , \quad \frac{f''}{s'^2 f} + \frac{2\mu}{\hbar^2}\left(\frac{E - V}{s'^2}\right) = \frac{\tilde{\sigma}}{\sigma^2} \tag{11}$$

The energy and potential terms in (11) may be decomposed in two pieces, which should provide a clear understanding for the contributions of $F$ and $f$ terms in (7) to the full solutions, such that $E - V = (E_F + E_f) - (V_F + V_f)$. Therefore, the second equality in Eq. (11) is transformed to a couple of equation

$$\frac{f''}{f} = \frac{2\mu}{\hbar^2}(V_f - E_f) \quad , \quad -\frac{\tilde{\sigma}}{\sigma^2}s'^2 = \frac{2\mu}{\hbar^2}(V_F - E_F) \quad , \tag{12}$$

where $f$ can be expressed in an explicit form considering the first part in (11) such as

$$f(r) \approx (s')^{-1/2} \exp\left[\frac{1}{2}\int^{s(r)}(\tau/\sigma)ds\right] . \tag{13}$$

## 3. APPLICATION

Up to now, this formalism has been used only to study exactly solvable systems [4, 6] and the related references therein. Therefore, it needs a meticulous modification to solve also quasi-exactly solvable systems as the one of interest in this article. To proceed, consider Eq. (5) where the potential terms are

$$V_{ES}(r) = V_F(r) + V_f(r) = \frac{1}{2}\mu w^2 r^2 + \frac{\hbar^2}{2\mu}\left(m^2 - \frac{1}{4}\right)\frac{1}{r^2} \tag{14}$$

which belongs to exactly solvable potential family having explicit expressions for the complete spectra, and the Coulomb interaction term

$$\Delta V(r) = \frac{e^2}{\varepsilon r} \tag{15}$$

that transforms the Schrödinger equation to the quasi-exactly solvable case [8]. The other term related to $L_z$ $(m\hbar w_c/2)$ can simply be considered as a shifting parameter that will automatically appear later in our energy expression.

Regarding that a quasi-exactly solvable potential behaves similar to the present consideration

$$V_{QES}(r) = V_{ES}(r) + \Delta V(r) \tag{16}$$

Eq. (7) in the preceding section needs to be rearranged as

$$\psi_{QES}(r) = \psi_{ES}(r)\Delta\psi(r) = [f(r)F(s(r))]g(r) \tag{17}$$

in which $g(r)$ is the modification function due to the presence of $\Delta V(r)$ term. The substitution of (17) in (6), where in this case $V \to V_{QES}$ and $E \to E_{ES} + \Delta E$, gives

$$F'' + \left(\frac{s''}{s'^2} + 2\frac{f'}{s'f}\right)F' + \left[\frac{f''}{s'^2 f} + \frac{2\mu}{\hbar^2}\left(\frac{E_{ES} - V_{ES}}{s'^2}\right)\right]F = 0 \tag{18}$$

which is Eq. (9), and one more equation, unlike exact solvability prescriptions,

$$\frac{g''}{g} + 2\frac{g'}{g}\left(\frac{f'}{f} + \frac{F's'}{F}\right) = \frac{2\mu}{\hbar^2}(\Delta V - \Delta E) \tag{19}$$

where the derivatives are taken with respect to $r$. Hence, the frame in (5) splits successfully into two parts to shed a light in revealing the inter-relation between exactly and quasi-exactly solvable potentials.

Evidently, the most significant piece in this model is Eq. (19) that is responsible for the calculation of corrections brought to the exact solutions obtained by the use of (18). Namely, the modifications because of the Coulomb interaction in (5) on the explicit solutions that belong to confining potential with the barrier term can be extracted with the consideration of (19). When $g(r)$ becomes constant, it can be easily seen that Eq (19) dies away and calculations reduce to the exactly solvable case, which provides us a testing ground for the reliability of the calculations.

We first deal with the closed algebraic solutions of the exactly solvable piece, Eq. (14), appearing in (5). This brief investigation opens a gate to the reader for the visualization of the explicit form of the center-of mass solutions without making any calculation. From the differential equation of the Laguerre polynomials [7], and considering Eqs.(10) and (18) together, one can see that

$$F(s) = e^{-s/2} s^{(\alpha+1)/2} L_n^\alpha(s) \quad , \quad \sigma = 1 \quad , \quad \tau = 0 \quad , \quad \widetilde{\sigma} = \frac{2n + \alpha + 1}{2s} + \frac{1 - \alpha^2}{4s^2} - \frac{1}{4} \quad , \quad n = 0,1,2,\ldots$$
(20)

Using the discussion presented in the previous section, in particular Eqs.(12) and (13), and accepting that $(s'^2/s) = a^2$ where $a$ is constant, we obtain

$$s = \frac{a^2 r^2}{4} \quad , \quad f = \sqrt{\frac{2}{a^2 r}}$$
(21)

leading to

$$V_F = \frac{\hbar^2}{2\mu} \left[ \frac{a^4 r^2}{16} + \frac{\alpha^2 - (1/4)}{r^2} \right] \quad , \quad E_F = \frac{\hbar^2 a^2}{4\mu}(2n + \alpha + 1) \quad , \quad V_f = \frac{\hbar^2}{2\mu}\left(\frac{3}{4r^2}\right) \quad , \quad E_f = 0$$
(22)

Clearly, sum of the two potential pieces $(V_F + V_f)$ should be equal to Eq. (14). This comparison yields that $\alpha = |m|$, which satisfies mathematical definitions in constructing Laguerre polynomials, and $a^2 = 4\mu w/\hbar$. Finally, the corresponding closed expressions for the energy and unnormalized wave functions are

$$E_{ES} = E_F + E_f = (2n + |m| + 1)\hbar w \quad , \quad \psi_{ES} = fF = r^{|m| + \frac{1}{2}} \exp(-w\mu r^2/2\hbar) L_n^{|m|}(w\mu r^2/\hbar) \quad (23)$$

which have the same form with those of the center-of mass solutions, keeping of course in mind that $r \to R$, $\mu \to M$ in this case. Note that the constant term $(m\hbar w_c/2)$ in (5) due to the magnetic field applied is now readily invoked to the solutions

$$E_{ES}(B) = E_{ES} + m\hbar w_c/2$$
(24)

in case, of course, $B \neq 0$.

After all, Eq. (19) can be expressed as

$$\frac{g''}{g} - 2\frac{g'}{g}\left\{\frac{1}{2r} + \left[\frac{a^2 r}{4} - \frac{1}{r}\left(2n + |m| + 1 - \frac{2(n+|m|)L_{n-1}^{|m|}}{L_n^{|m|}}\right)\right]\right\} = \frac{2\mu}{\hbar^2}\left(\frac{e^2}{\varepsilon r} - \Delta E\right) \quad (25)$$

as $\left(L_n^{|m|}\right)' = (n/s)L_n^{|m|} - ((n+|m|)/s)L_{n-1}^{|m|}$. This vital part of the formalism is a kind of Riccati equation. The exhaustive analysis of Eq. (25), in the light of related literature, guides us to choose the correction term as

$$g(r) = 1 + \sum_j \beta_j r^j \quad , \quad j = 1,2,3,\ldots \quad (26)$$

that modifies the solutions in (23) for small $r$ – values because of the natural existence of the Coulomb term in the total Hamiltonian. In the above definition, $\beta_{j=1}$ is responsible for constructing the potential term in (25) having different structure for each $j$ – value in order to keep the term related to the Coulomb interaction unchanged with the increasing degree of $g(r)$,

$$\frac{2\mu}{\hbar^2}\left(\frac{e^2}{\varepsilon r}\right) = \left(\frac{1+2|m|}{r}\right)\beta_1 \quad (27)$$

For a clear understanding, let us consider first $n = 0$ and $j = 1$ case for which $L_{-1}^{|m|} = 0$ in Eq. (25) and subsequently

$$\beta_1 = \pm\left(\frac{a}{\sqrt{2+4|m|}}\right) = \pm\sqrt{\frac{2\mu w}{(1+2|m|)\hbar}} \quad (28)$$

As we deal with the two-electron interaction, the positive root should be chosen. This proper chose produces

$$\frac{2\mu}{\hbar^2}\Delta E_{j=1} = \frac{a^2}{2} \Rightarrow \Delta E_{j=1} = \hbar w \quad , \quad \Delta\psi = g_{j=1} = 1 + \frac{\sqrt{2\mu w}}{\sqrt{(1+2|m|)\hbar}} r \quad (29)$$

Similarly, in case of $j = 2$ while again $n = 0$, the form of the modification function becomes $g(r) = 1 + \beta_1 r + \beta_2 r^2$ that forced calculations to reproduce now three roots for $\beta_1$

$$\beta_1 = 0, \pm\sqrt{\frac{4\mu(3+4|m|)w}{(1+2|m|)^2 \hbar}} \quad (30)$$

depending upon obviously the appropriate choice of $\beta_2 \left(= 2\mu w/\hbar(1+2|m|)\right)$. This simply can be understood as a kind of compensation to be able to validate Eq. (27), since $g(r)$ now is the

second order polynomial. Again, the physically meaningful root should be the positive one as the others do not satisfy Eq. (27), at least for the present consideration. It is however reminded that if one interests, unlike our case, in two-body interaction having opposite charges, negative $\beta_1$ values should of course be chosen. Proceeding with the use of convenient $\beta_1$ value for the present consideration then one gets

$$\Delta E_{j=2} = 2\hbar w \quad , \quad \Delta \psi = g_{j=2} = 1 + \sqrt{\frac{4\mu(3+4|m|)w}{(1+2|m|)^2 \hbar}} r + \frac{2\mu w}{\hbar(1+2|m|)} r^2 \qquad (31)$$

This procedure is well adapted to the use of software systems such as Mathematica and allows the computation to be carried out up to high orders of the polynomial in (26). For any given $j$-value, simple algebraic manipulations provide a clean route in understanding the interconnection between the node numbers $(n)$ in the wave function and orders of $g(r)$. The increase in the value of $j$ for different radial quantum numbers $(n)$ does not imply special difficulty since the node number of the total wave function in (17) is merely defined by the structure of $g(r)$. Our careful calculations have nicely revealed that $2n+1 \leq j$, consequently $n \leq (j-1)/2$ being one of the important observations in the present study. Two examples above justify this fact. To clarify this point, a small piece of the analytical results obtained $(n = 0,1)$ are illustrated in Table 1.

The other significant observation encountered through the work discussed in this article is the relation between $j$ and the number possible roots for $\beta_1$. More explicitly, one should find $j+1$ roots for $\beta_1$ if the degree of $g(r)$ is $j$. From these mathematically possible roots, the physically acceptable ones corresponding to the nodes can easily be picked up by $n \leq (j-1)/2$. For instance, we have 3 roots in (30) if $j = 2$, but from the physics point of view we are forced to choose one of them due its positive sign. This is indeed governed by $n \leq (j-1)/2$ representing a concrete relation that enables us to decide precisely regarding the structure of the wave function we deal with. Thus, in case of $j = 2$, there should be only one $\beta_1$ value which certainly should produce a wave function without any node, since $n = 0$. Further, remembering the well known connection between the principal quantum number $(n_p = 1,2,...)$ related to energy levels and the radial quantum number $(n)$, $n_p = n + |m| + 1$ in two-dimensions, one can easily determine the level of a state function for arbitrary azimuthal quantum numbers.

The final outcome of the calculations comes from the attentive investigation of Eq. (27). For simplicity, let us concentrate on the lowest case where $n = 0$ and $j = 1$ for which $\beta_1$ is given by (28). By substituting this value in (27), and keeping in mind that the left hand side of the equation should remain unchanged at each order of $j$, one can find that

$$w = \frac{2\mu e^4}{\varepsilon^2 (1 + 2|m|)\hbar^3} \tag{32}$$

From the section 2 we know that $w = [w_0^2 + (w_c^2/4)]^{1/2}$, hence the cyclotron frequency takes the form

$$w_c \equiv \frac{(e/2)B}{\mu c} = \frac{eB}{m^*c} = 2\left[\left(\frac{m^* e^4}{\varepsilon^2 (1 + 2|m|)\hbar^3}\right)^2 - w_0^2\right]^{1/2} \tag{33}$$

This feature implies that for specific values of the magnetic field, the Hamiltonian in (5) can be solved exactly. In other words, the Coulomb interaction in (5) destroys the general symmetry, reducing the problem to the non-exactly solvable case, nevertheless the magnetic field can restore the symmetries again for its particularly chosen values. In connection with this it is remarkable that the novel prescription in (19), and its extended form Eq.(25), works out the two-electron correlation problem existed in (5) at once, in an elegant manner, bypassing the difficulties and cumbersome calculations involved through the three-term recursion relations and group theoretical approach used in Refs. [2,8,9,10]. Moreover, for the limiting case where $B = 0$, the choice of a special characteristic length $(\ell_0)$ for the quantum dot maintains reproducing closed analytical expressions for the relative motion of electrons in the device, as $w_0 = \hbar/m^* \ell_0^2$. Additionally, from Eq. (33), when the magnetic field increases, apparently the effective frequency $(w)$ gains larger values, the dot size $(\ell_0)$ decreases. From Table 1, it is evident that the smallest frequency has zero node while the second largest one has one node, and so on.

In finalizing, we combine the above results with Eqs.(23) and(24)

$$E_{rm} = (j + 2n + |m| + 1)\hbar w + \frac{m\hbar w_c}{2} \quad , \quad \psi_{rm} = g_j(fF) \quad , \quad j = 1,2,3\ldots \tag{34}$$

to define the internal motion of the electrons algebraically. For the complete consideration, it is reminded however that $E = E_{cm} + E_{rm}$ and $\Psi = \Phi(r)\Theta(R)$ where the connection between $\Phi$ and $\psi$ has been shown by (4). The results obtained and the observations discussed above support the findings in [2,3,8,9,10] concerning with the same problem within the frame of alternative treatments.

Further, from (34), the effect of the electron correlation connecting to the $j$ – value on the whole energy spectra of the quantum dot is now explicitly attainable. We see that the magnetic field shifts the ground state spectrum with $n = 0$, $m = 0$ to those of higher angular momenta $(n = 0, m \succ 0)$ in order to decrease the Coulomb electron-electron repulsion, as reported in [11]. This can be seen from a careful investigation of Eqs. (27,28,30,32,33), which are eventually employed by (34), and the physics behind it is understood as follows. As $B$ increases, the electrons are further squeezed in the quantum dot, resulting in an increase of the repulsive Coulomb energy between electrons, and in effect the energy levels. In connection with this, the increase in the magnetic field strength causes to the increase in the energy of the state $m = 0$ while the energy of the states with $m \succ 0$ decreases. This leads to a sequence of different ground states. The behaviour for excited levels seems similar, see Table 1.

Finally, through the discussion in this section, it is stressed that we have started with a natural consideration that Eq. (14) represents an exactly solvable piece in (5). Instead, on the contrary, one may start with an alternative approach to the same problem, namely the Coulomb potential with the barrier term may be considered as the exactly solvable part in (5) and all the procedure is then carried out for the new perturbing potential $(\Delta V = \mu w^2 / 2 r^2)$. This significant consideration, which does not cause any physical problem, can be used to check the reliability of the calculation results obtained,

$$E_{ES}^{HO} + \Delta E_C = E_{ES}^C + \Delta E_{HO} \quad , \quad \psi_{ES}^{HO} \Delta \psi_C = \psi_{ES}^C \Delta \psi_{HO} \tag{34}$$

producing indeed exactly identical algebraic expressions.

As concluding remark, due to its simplicity and accuracy in particular for small orders of the polynomial at low-lying states we believe this method to be competitive with other techniques developed to deal with the problem under consideration. As a matter of fact, the wide applicability of the scheme used can be readily observed by raising the correction order, $j-$value, a step which does not in principle bear any technical difficulty.

## 4. CONCLUDING REMARKS

Summarizing, we have calculated the discrete energy spectra for two electrons in a two-dimensional harmonic well that serves as a simple but suitable model for quantum dots on semiconductor interfaces. We have shown that exact solutions to the Schrödinger equation for potentials of the form Coulomb plus harmonic oscillator can be found subject to a constraint on the ratio between the strengths of these potential terms. This means that the symmetries of the Hamiltonian for such systems can be recovered for particularly chosen values of the magnetic field and the geometric size of the dot. The most appealing feature of quantum dots as compared to other atomic-like systems like donors in semiconductors is the tunabilty of their size and electron number by technological means. Taking this point of view, it would be interesting to extend the present scenario, which has proven its success for the simplest quantum dot, to other more complex systems. In particular, the present results would be useful in perturbational treatments of the exact spectra of a few particle systems, and thus provide a further insight on discussion of the fractional nature of such systems. Along this line, the works are in progress.

| $j$ | $n$ | $\Delta E$ | $\beta_1$ (only the positive roots have been used) | $(w_c/2) = \sqrt{w^2 - w_0^2}$ |
|---|---|---|---|---|
| 1 | 0 | $\hbar w$ | $\pm \sqrt{\dfrac{m^* w}{(1+2|m|)\hbar}}$ | $\sqrt{\left(\dfrac{m^* e^4}{(1+2|m|)\varepsilon^2 \hbar^3}\right)^2 - w_0^2}$ |
| 2 | 0 | $2\hbar w$ | $0, \pm \sqrt{\dfrac{2m^*(3+4|m|)w}{(1+2|m|)^2 \hbar}}$ | $\sqrt{\left(\dfrac{m^* e^4}{2(3+4|m|)\varepsilon^2 \hbar^3}\right)^2 - w_0^2}$ |
| 3 | 0 | $3\hbar w$ | $\pm \sqrt{\dfrac{10+10|m|+\sqrt{73+64|m|(2+|m|)}\, m^* w}{\hbar(1+2|m|)^2}}$ | $\sqrt{\left(\dfrac{m^* e^4}{\left(10+10|m|+\sqrt{73+64|m|(2+|m|)}\right)\varepsilon^2 \hbar^3}\right)^2 - w_0^2}$ |
| 3 | 1 | $3\hbar w$ | $\pm \sqrt{\dfrac{10+10|m|-\sqrt{73+64|m|(2+|m|)}\, m^* w}{\hbar(1+2|m|)^2}}$ | $\sqrt{\left(\dfrac{m^* e^4}{\left(10+10|m|-\sqrt{73+64|m|(2+|m|)}\right)\varepsilon^2 \hbar^3}\right)^2 - w_0^2}$ |
| 4 | 0 | $4\hbar w$ | $0, \pm \sqrt{\dfrac{25+20|m|+3\sqrt{33+8|m|(5+2|m|)}\, m^* w}{\hbar(1+2|m|)^2}}$ | $\sqrt{\left(\dfrac{m^* e^4}{\left(25+20|m|+3\sqrt{33+8|m|(5+2|m|)}\right)\varepsilon^2 \hbar^3}\right)^2 - w_0^2}$ |
| 4 | 1 | $4\hbar w$ | $\pm \sqrt{\dfrac{25+20|m|-3\sqrt{33+8|m|(5+2|m|)}\, m^* w}{\hbar(1+2|m|)^2}}$ | $\sqrt{\left(\dfrac{m^* e^4}{\left(25+20|m|-3\sqrt{33+8|m|(5+2|m|)}\right)\varepsilon^2 \hbar^3}\right)^2 - w_0^2}$ |

**Table 1**. Low-lying quantum state energy corrections, potential parameters in Eq. (27) and cyclotron frequencies at each successive orders of $g(r)$, which are required by (33). In the table, $m^*$ and $|m|$ represents the effective mass of the electrons and arbitrary azimuthal quantum numbers, respectively.